# Mapping and Comparing Data Governance Frameworks

A benchmarking exercise to inform global data governance deliberations

*Sara Marcucci, Natalia González Alarcón, Stefaan G. Verhulst, and Elena Wüllhorst*

*February, 2023*

## Abstract

Data has become a critical resource for organizations and society. Yet, it is not always as valuable as it could be since there is no well-defined approach to managing and using it. This article explores the increasing importance of global data governance due to the rapid growth of data and the need for responsible data use and protection. While historically associated with private organizational governance, data governance has evolved to include governmental and institutional bodies. However, the lack of a global consensus and fragmentation in policies and practices pose challenges to the development of a common framework.

The purpose of this report is to compare approaches and identify patterns in the emergent and fragmented data governance ecosystem within sectors close to the international development field, ultimately presenting key takeaways and reflections on when and why a global data governance framework[1] may be needed. Overall, the report highlights the need for a more holistic, coordinated transnational approach to data governance to manage the global flow of data responsibly and for the public interest.

The article begins by giving an overview of the current fragmented data governance ecology, to then proceed to illustrate the methodology used. Subsequently, the paper illustrates the most relevant findings stemming from the research. These are organized according to six key elements: (a) purpose, (b) principles, (c) anchoring documents, (d) data description and lifecycle, (e) processes, and (f) practices. Finally, the article closes with a series of key takeaways and final reflections.

[1] In this report, "framework" refers to any type of document or project subject to being analyzed under this analysis template.

# 1. Introduction: Identifying Patterns in a Fragmented Data Governance Ecology

Data has become a global asset; therefore, how it is managed and governed has become a priority for a diverse number of stakeholders around the world. Historically, data governance has often been associated with private organizational or corporate governance approaches. However, as technological innovation and the amount of data have increased rapidly in recent years, the notion of data governance has evolved to include governmental and institutional bodies. There have been increasing calls for a global data governance framework that would help manage the global flow of data responsibly, while ensuring a necessary balance between its undeniable potential and equally undeniable risks. The World Development Report 2021 by the World Bank, for instance, acknowledges the increasing development of data governance arrangements, yet incipient, and alerts on how the current regulatory efforts might be inadequate for the 'majority world' (Mungai et al. 2022). There are different reasons associated, some related to significant gaps in infrastructure, security, institutional and safeguard mechanisms, and others linked to countries' unlike needs and priorities. "A global consensus would give individuals and enterprises confidence that data relevant to them carry similar protections and obligations no matter where they are collected or used. (..) It would also establish ground rules for the exchange of data between commercial use and the public good" (World Bank 2021, 297).

The search for a global data governance framework emerges from a complex landscape that bridges policy and practice, and encompasses a number of different domains, such as data management, data ethics, and data protection. "The approach to governing data and data flows varies considerably among the major players in the digital economy, and there is little consensus at the international and regional levels" (UNCTAD 2021, 98). In addition, different frameworks within and across countries, regions, sectors, and organizations have resulted in a patchwork of policies, frameworks and practices, leading to a fragmented ecology that poses certain challenges to the evolution of a common framework.

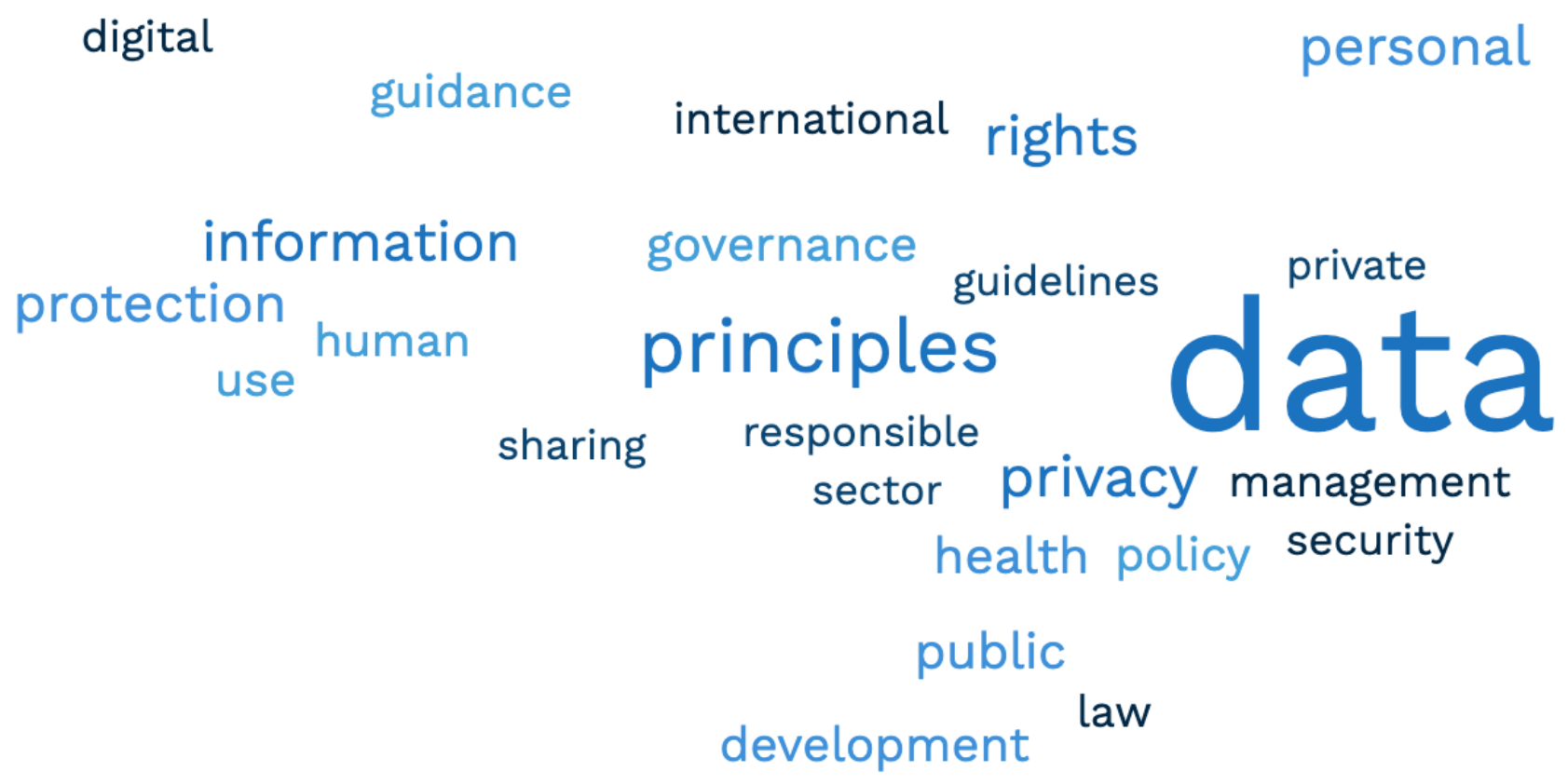

*Figure 1: Word cloud of key words emerging from the frameworks analyzed in this report.*[2]

This fragmentation is heightened by the dynamism of the ecology, with new solutions, often technical in nature, being released on a regular basis, often seeking to update legacy approaches that are no longer fit for the challenges and opportunities of new data realities. In addition, emerging technologies and concepts, for instance Artificial Intelligence or Distributed Ledger Technologies, lead to new governance needs and result in sometimes ad-hoc extensions to existing data governance approaches (World Bank 2021, 271; Anthony 2022, 293).

While some of the frameworks–such as data protection regimes–are more mature and lend themselves to standardization and codification, others fall short, representing more reactive approaches. As a result, efforts to harmonize and coordinate the various frameworks have often been led by business or professional associations, standard-setting bodies, or international fora connecting national data protection authorities rather than by any entity with truly global reach and credibility. Because of the diversity of entry points, concerns, and interests, there is a wide diversity of actors advocating for different approaches, often operating in silos without much engagement or coordination. However, the increasing interconnection and interdependence within the global data economy urge to evolve towards a more holistic, coordinated transnational approach that might require new and innovative global governance  (UNCTAD 2021, 215).

[2] The word cloud was generated from a short description of each framework, which is included in the repository. The descriptive data was cleaned through removing words such as yes, verbs, commas, and quotation marks. 25 keywords of the cleaned descriptive data were visualized using the word cloud generator freewordcloudgenerator.com. Each term shown in Figure 1 occurs at least 10 times.

## 2. Why a Global Governance Framework May Be Necessary

As mentioned above, this paper seeks to inform current deliberations on whether a global data governance framework is needed. The following aims to summarize some of the existing literature exploring why and when global governance may be beneficial, focusing particularly on (a) global coordination to prevent harmful fragmentation, (b) the advancement of global principles and values, and (c) using data as a resource to advance global public goods.

### 2.1. Global Coordination to Prevent Harmful Fragmentation

As the world becomes increasingly interconnected, it is ever more urgent to build systems of cooperation that allow multiple and diverse actors to collaborate and make use of a dedicated framework for sharing information, expertise, and experience. It could seem particularly crucial to develop a standardized approach to data in certain sectors or at certain moments, such as in times of humanitarian crisis, as this will ultimately enable more coordination and prevent harmful fragmentation. Given the multiple and complex interactions between regulations and asymmetries at local, national, and international levels, fragmentation may have profound implications on individuals and businesses, both intended and unintended, for virtually all aspects of our daily lives (Fay 2022).

Further, the absence of a systematic global approach to data governance may create inequitable consequences for low- and middle-income countries as it would be harder for them to participate in the global digital economy and develop their own frameworks responsibly (Pisa and Nwankwo 2021). Even more ambitious approaches argue that a "new Bretton Woods-style agreement" is necessary to redefine the global governance model in a digital and hyper-globalized world (Medhora and Owen 2020). By enhancing cooperation and shared standards and principles, global data governance might allow nations and organizations to collectively take advantage of the potential data harbors to face common challenges and respond to collective needs.

### 2.2. Advancing Global Principles and Values

A global data governance framework would enable international cooperation and coordination to promote globally shared principles and values, such as human rights, and to further anchoring frameworks such as the Universal Declaration of Human Rights or the SDGs. In fact, data is playing an increasingly important role in the humanitarian field. From the United Nations to private Non-Governmental Organizations (NGOs), organizations across the world are starting to adopt data on ever larger scales to enable more agile, efficient and evidence-based decision-making to promote human rights and other global values.

For example, the World Economic Forum (WEF) has proposed a new data governance model called Authorized Public Purpose Access (APPA), defined as "a model for realizing value by permitting access to data for specific, agreed public purposes (...) through processes that do not rely exclusively on explicit, individual consent as a means of protecting human rights" (World Economic Forum 2021). The ultimate goal is guaranteeing individual human rights regarding data use, not limited to privacy rights. Also, as part of the 2030 Agenda, the UN reaffirmed its commitment to international law and emphasized that all efforts shall be implemented in a manner that is consistent with human rights (United Nations High Commissioner for Human Rights 2018). Given the international and often borderless nature of these objectives, and considering the key role data has in achieving them, a global data governance framework is essential.

### 2.3. Using Data as a Common Resource to Advance Global Public Goods

Increasingly, data is a resource for supporting global public goods. It is crucial that no single entity has sovereignty over data and that these collective public goods are managed responsibly, in a manner that secures and preserves them for all of humankind. Lately, there has been a debate on treating data as 'commons' when the economic characteristics of data define it as an intangible non-rival asset, which may suggest an opposite definition. However, it is a practical approach when considering data governance and regulation.

Ostrom's principles (Ostrom 2012) for managing shared resources (commons) offer insightful guidelines to reach an agreement about rules for accessing data when some individuals may need to sacrifice personal benefit for the greater common good (Coyle et al. 2020). The principles help understand the asymmetries of information and incomplete agreements that characterize the data economy and provide innovative ways to govern shared resources. For example, defining the rights of different entities to control, access, use and share data, and establishing monitoring and auditing mechanisms for data use and sharing (Coyle et al. 2020) might help to address the governance challenge of preventing misuse of sensitive data while fostering the reuse of data to create social value and potential 'knowledge spillovers' (Coyle 2020) (Aaronson 2022).

## 3. Methodology

The findings and recommendations of this article are based both on empirical data as well as an analytical framework, which permits to derive broader conclusions from the data.

### 3.1. Empirical Data: Sampling Strategy

In conducting this research, more than 100 data governance documents were identified, of which 58 were curated and surveyed in detail, across 37 organizations (non-governmental, intergovernmental and independent), 8 national or local government entities and 4 regional bodies. Appendix A includes the full list of the frameworks analyzed in depth, while the Template of Analysis in Section 3.3. provides a detailed assessment strategy of the frameworks considered.

This article was first initiated as a report to inform the United Nations High-level Committee on Programmes (HLCP) and has subsequently evolved to provide insights to other intergovernmental and governmental institutions, non-governmental organizations, academic and research institutions, and other stakeholders that aim to use and govern data.

The sampling strategy took into account the following considerations:

- **Timeframe:** To ensure relevance and validity, this research focused on frameworks created within the last ten years (2013-2022).

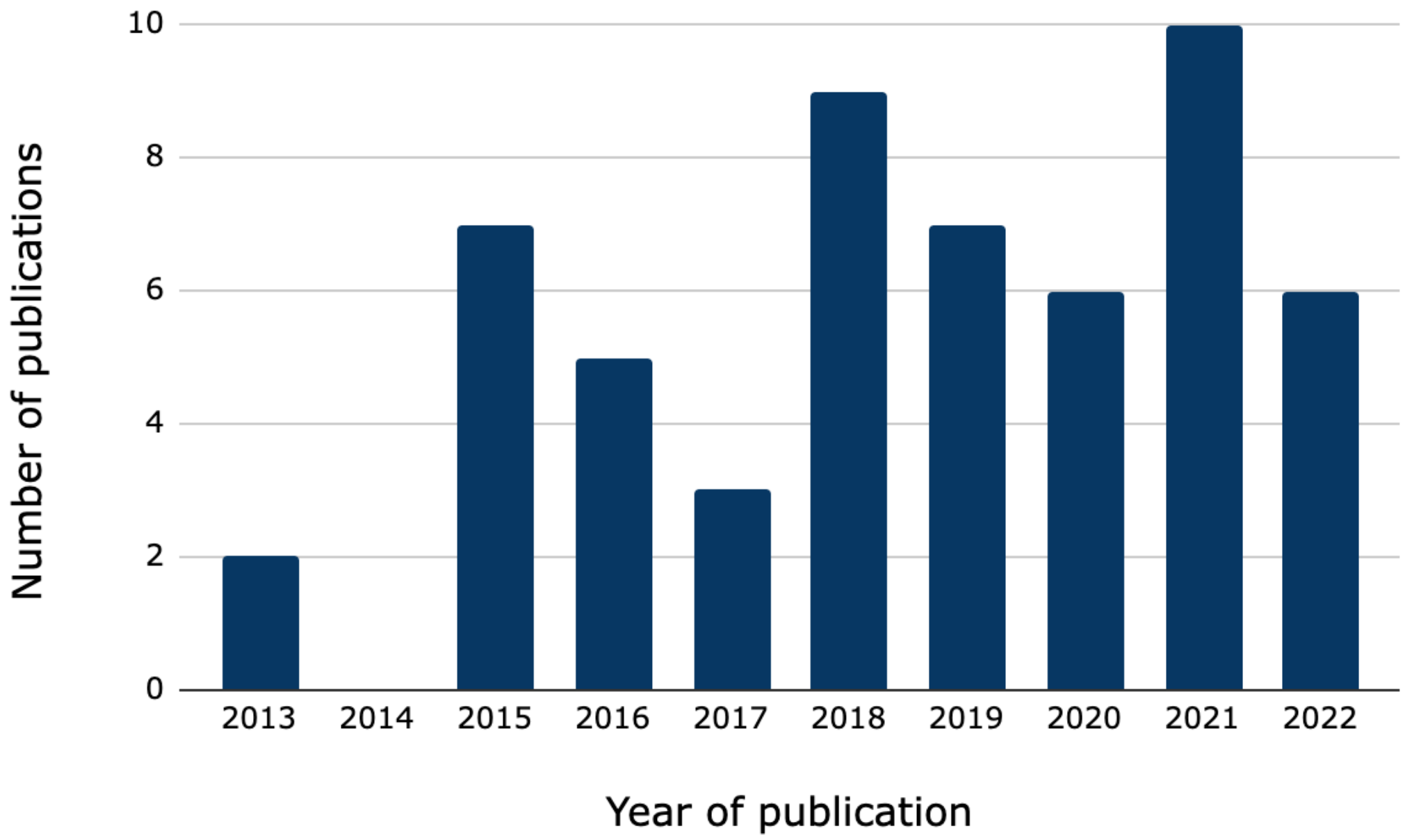

*Figure 2: Number of publications per year.*

- **Variety of frameworks:** This research aims to represent the variety and heterogeneity of existing data governance frameworks and approaches, and thus includes a wide range of examples and formats, encompassing principles, guidelines, and regulatory frameworks and standards.

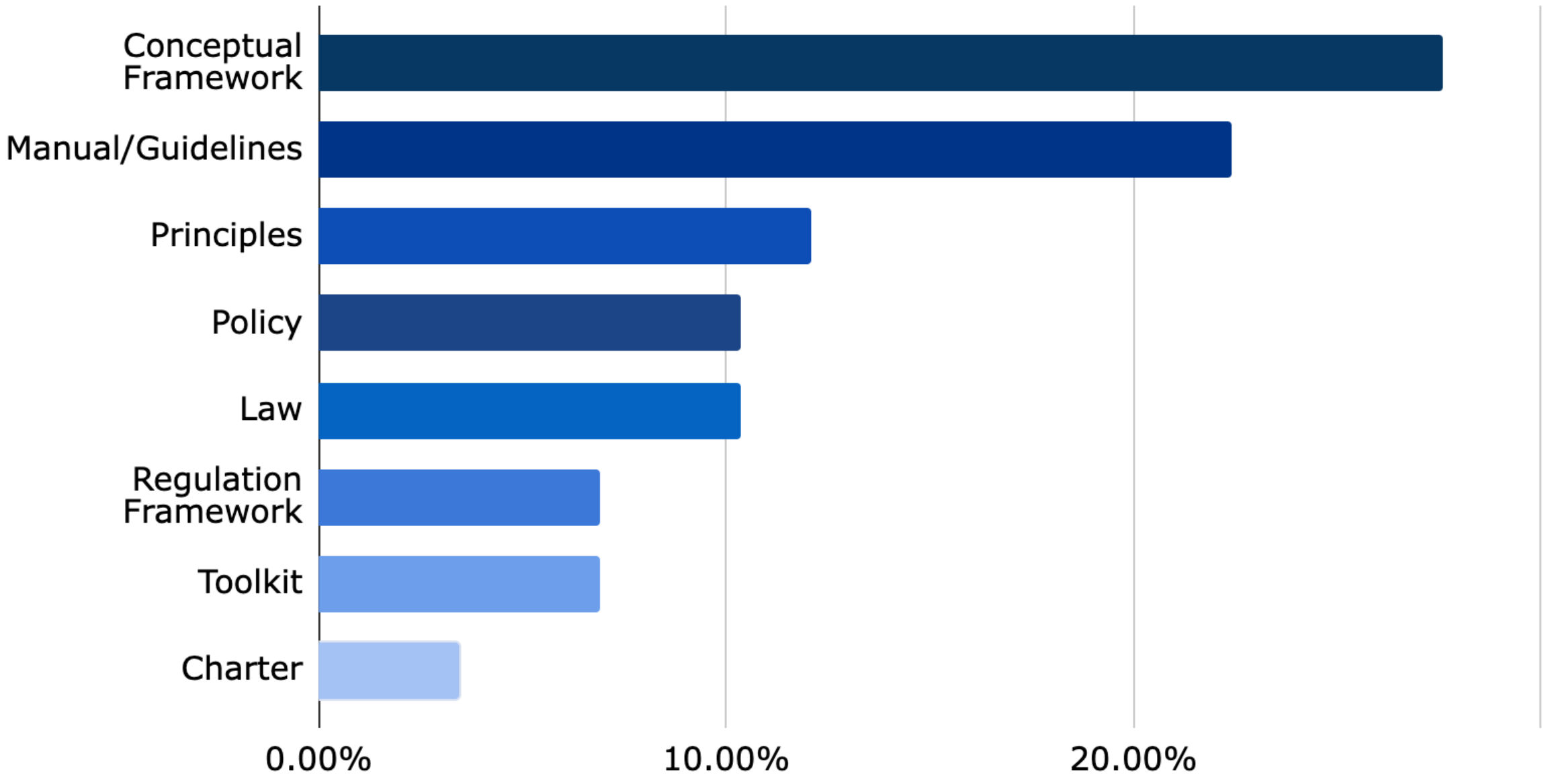

*Figure 3: Types of frameworks.*

- **Types of organizations:** The research prioritized public sector efforts over private-sector-oriented ones, cognizant of how the public sector plays a critical role in setting policies and regulations for data governance. While private sector efforts are also important in this regard, to maintain the scope of the research manageable, they were not the primary focus of this research. Thus, the sample prioritizes governmental and intergovernmental institutions, non-governmental organizations, academic and research institutions, and international independent (sectoral) coalitions.

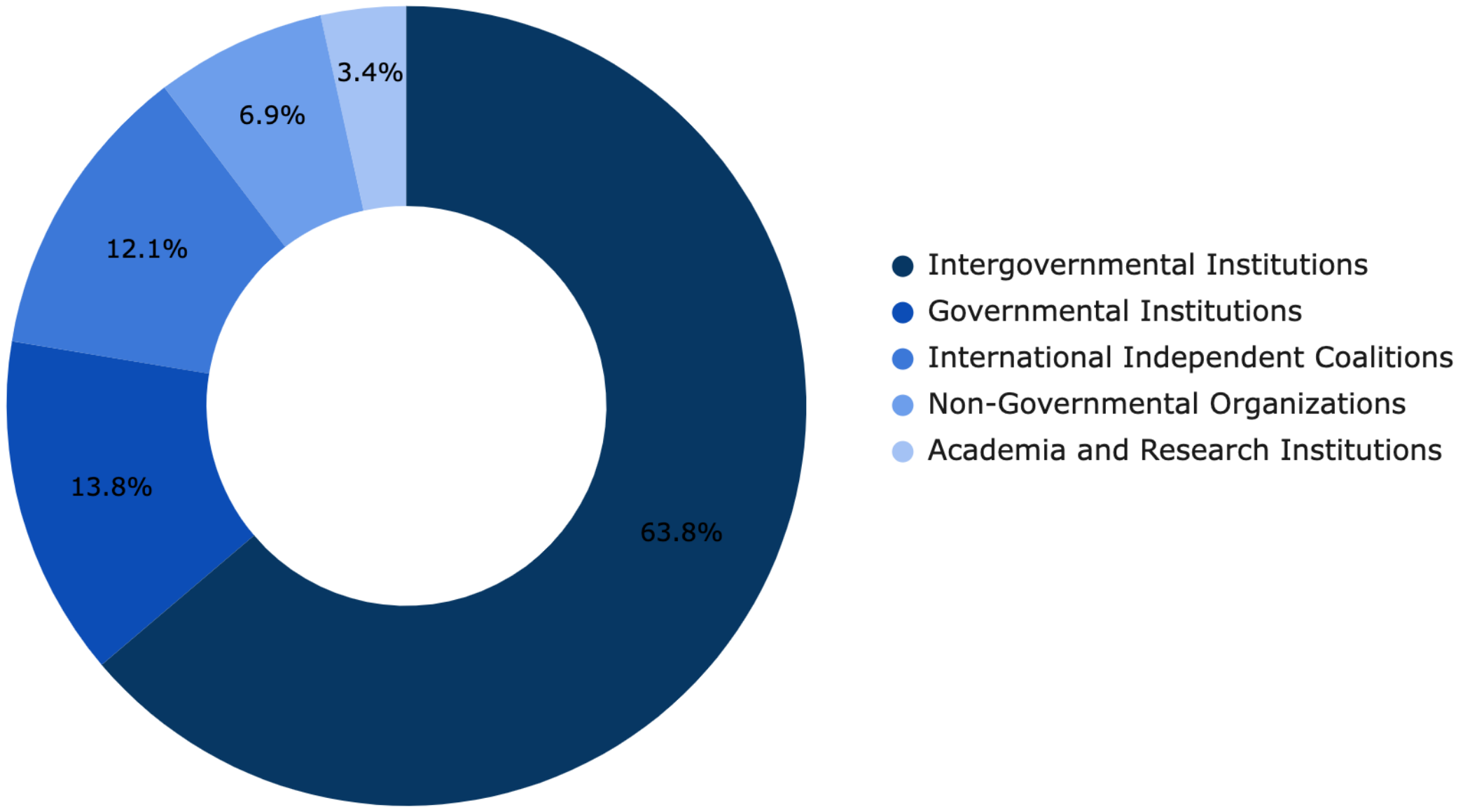

*Figure 4: Type of organization.*

- **Geographical scope:** When considering the geographical scope, this research aims to cover various levels of jurisdictions, including global, regional, and national frameworks. Although diversity and comprehensive representation across regions was sought, a stocktaking of various data governance frameworks and practices from the Global South is still largely missing. On the one hand, that is due to a lack of a clear and widely spread understanding of the frontier of data governance best practices. This might limit some governments from finding a suitable reference when building and prioritizing their own (Chen 2021). On the other hand, sometimes legal and regulatory frameworks for data are inadequate in lower-income countries, which too often face substantial gaps in safeguards and shortages of infrastructure. Indeed, as the World Development Report 2021 by the World Bank notes, "less than 20 percent of low- and middle-income countries have modern data infrastructure such as colocation data centers and direct access to cloud computing facilities. Even where nascent data systems and governance frameworks exist, a lack of institutions with the requisite administrative capacity, decision-making autonomy, and financial resources holds back their effective implementation and enforcement" (World Bank 2021, 13).

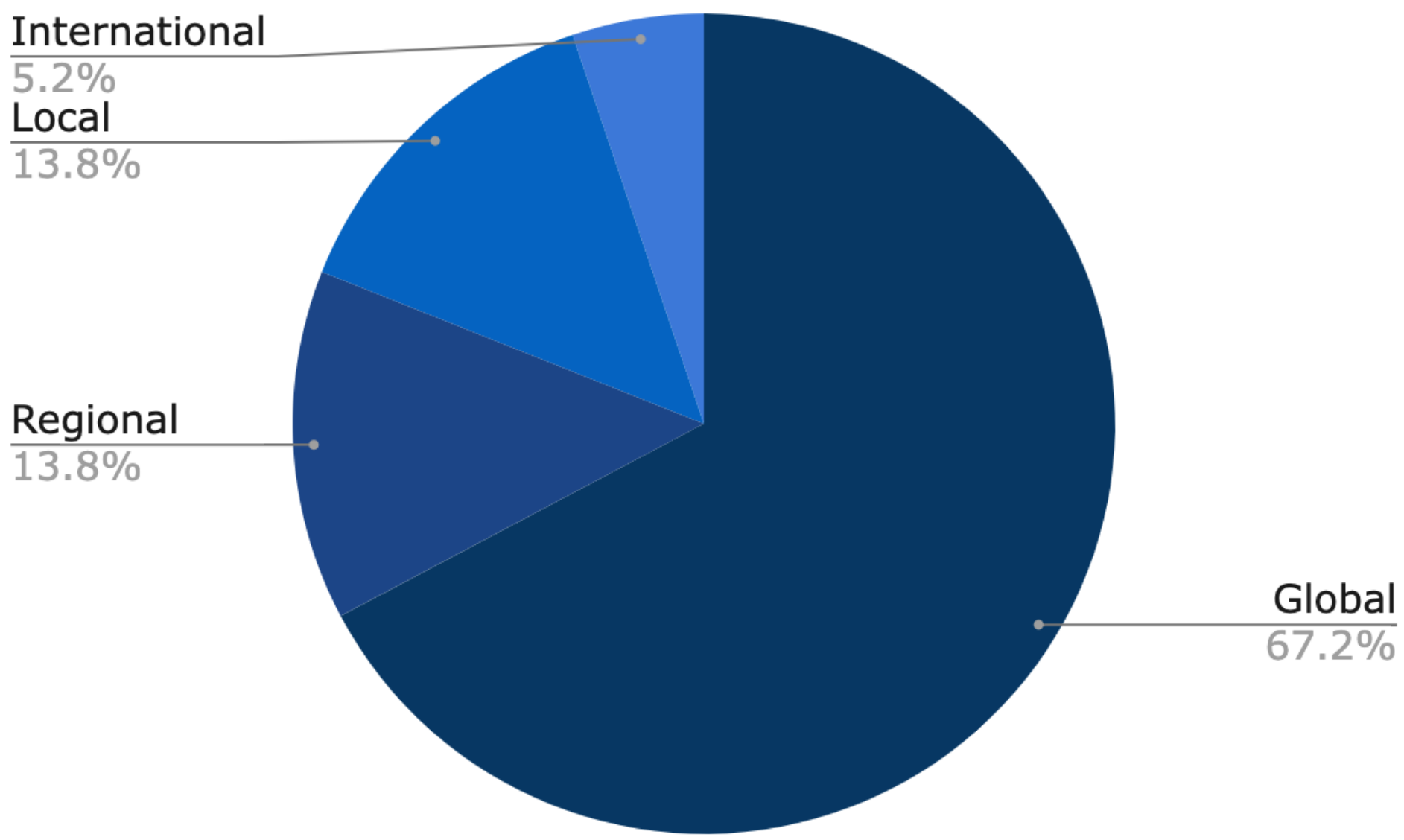

*Figure 5: Geographical scope.*

- **Sectoral diversity:** As mentioned previously, this research was initially conducted to inform the UN HLCP Group, which is primarily focused on issues related to the UN's programs and strategies for sustainable development. Given the Group's focus on humanitarian and development issues, the research prioritized frameworks and policies that were most relevant to those areas.

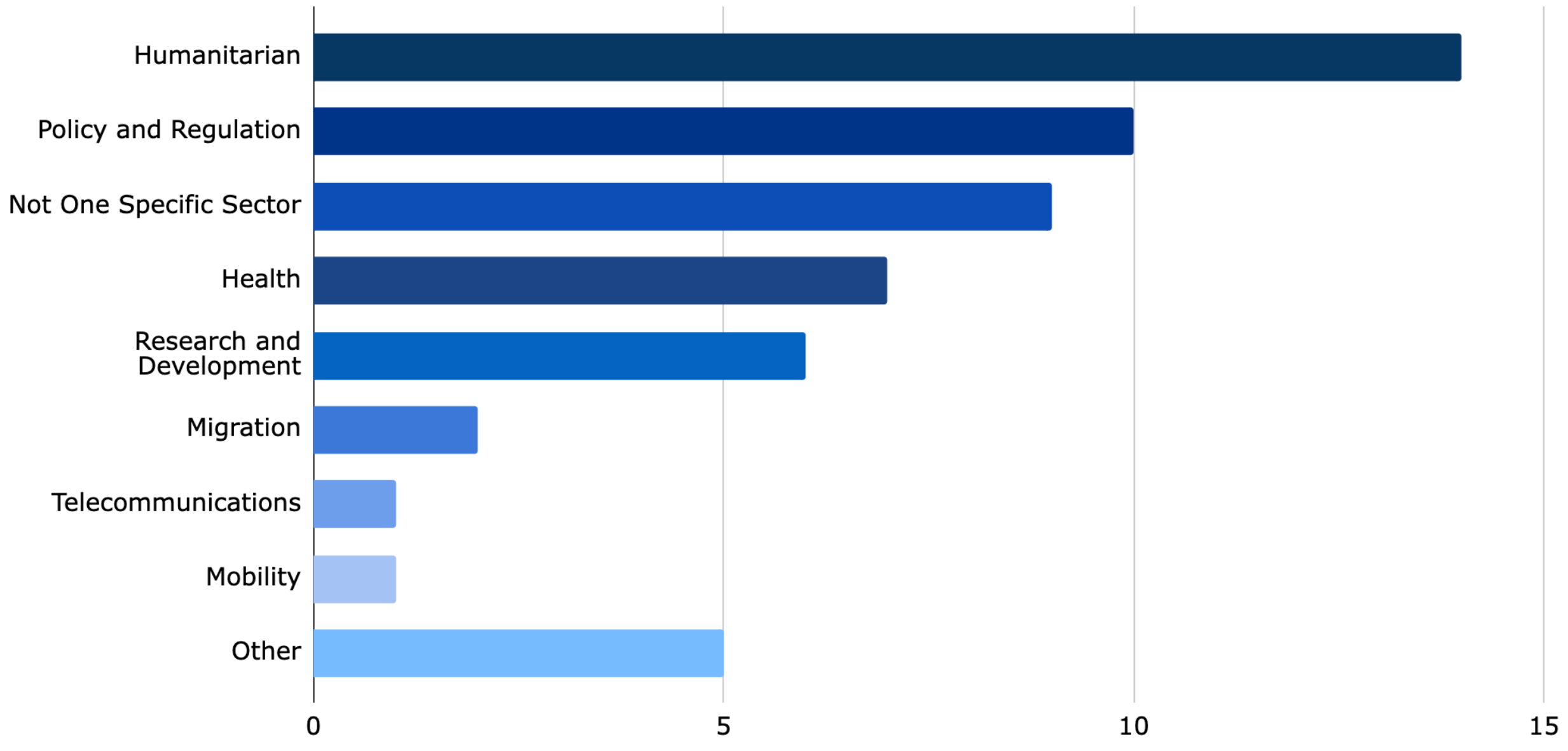

*Figure 6: Sectoral diversity.*

## 3.2. UN Agencies Sub-sample

Given the initial intent to inform UN decisions on global data governance, it is worth declaring the weight the UN-system agencies represent in the sample. Of the frameworks analyzed, twenty-six correspond to UN-system data governance frameworks. Twenty-five of those have a global scope except for the Pan American Health Organization's (PAHO) National Data Governance Framework, since PAHO serves as the Regional Office for the Americas of the World Health Organization. Of the UN Agencies sub-sample 69% are conceptual frameworks or guidelines. And, 42% are in the humanitarian sector.

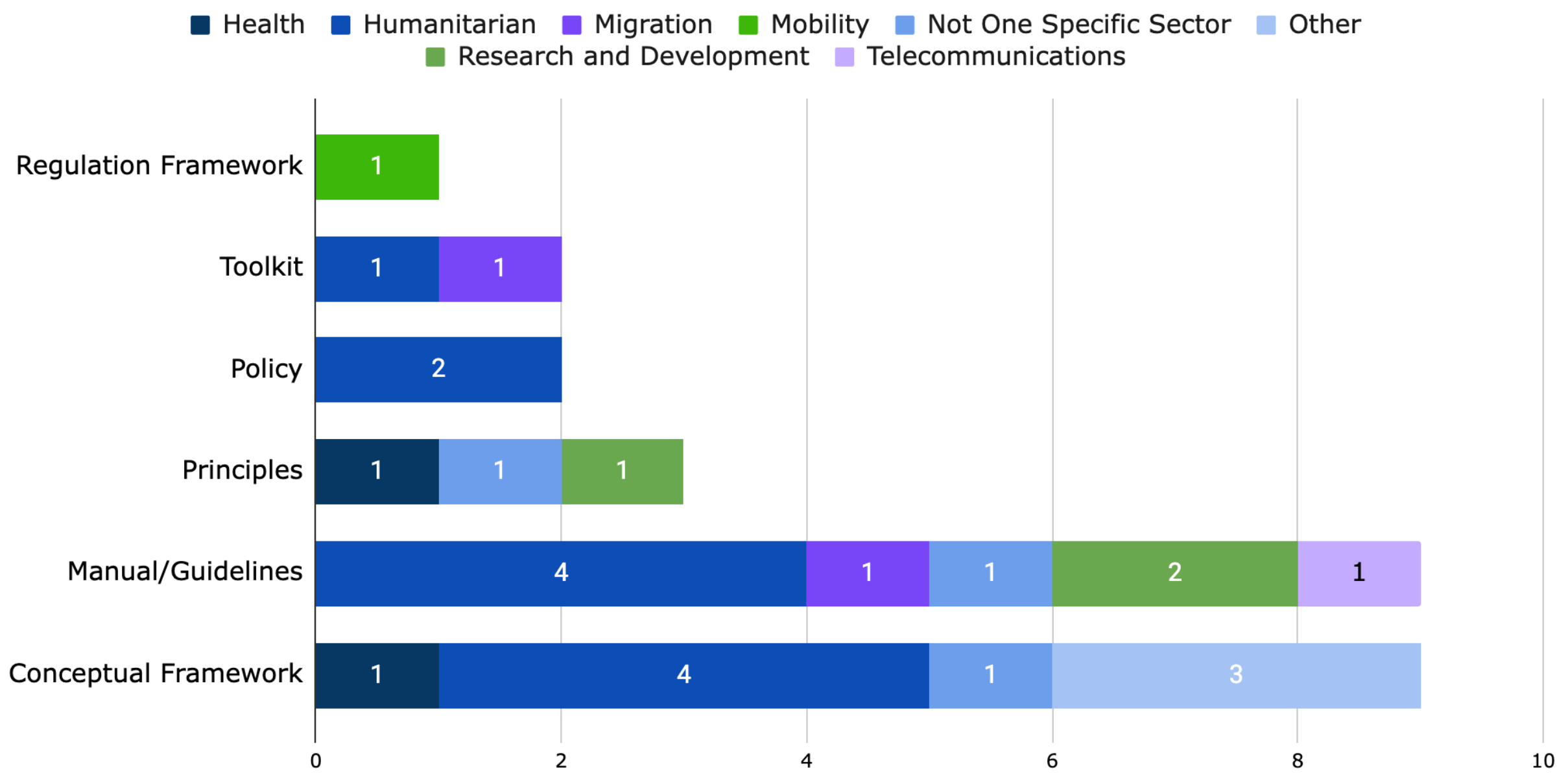

*Figure 7: Types and sectors of UN-system data governance frameworks.*

## 3.3. Analytical Framework

Once the empirical sample was assembled, the frameworks were analyzed through a conceptual prism consisting of the following "Template of Analysis".

| |
|---|
| **Purpose:** Purpose serves as the guiding objective of data governance frameworks. It illustrates the reason why a framework is needed in the first place, identifying a gap, and indicates the value the framework wants to bring about by filling that gap. |
| *Purpose:* Does the governance framework clarify its goals and objectives? |
| **Principles:** Principles serve as the guidance for a governance framework, ensuring that all activities are aligned with specific commonly agreed criteria and allows for easier interpretation. |
| *Principles:* Are there principles to guide the framework, and what are they? |
| *Anchoring:* Is the legal basis or other anchor documentation upon which the policies and principles built sufficiently explained? What is the nature of that basis? |
| **Data**: Describing and defining the data handled by the organization facilitates the understanding of the framework and pushes the organization to determine and justify the data they access and use. It also helps identify the data they seek to govern within the data value chain/ data lifecycle. |
| *Data Description:* Do the frameworks define the data they oversee? Is it personally identifiable data or not? |
| *Data Lifecycle:* Does the framework describe the value-chain of data and the benefits and risks at each stage? (e.g., data localization) |
| **Processes and Practices**: Operationalize the framework and ensure the principles are supported, and the processes are undertaken and monitored. |
| *Governance Roles:* Does the framework explain any roles and functions that are tasked with the implementation of the framework? (e.g., code of conduct, data sharing agreement) |
| *Tools*: What tools and practices are specified to implement the framework? |
| *Monitoring and Evaluating:* What monitoring and evaluation mechanisms are implemented? |

## 3.4. Limitations

Despite the comprehensive nature of this research, there are several limitations to consider.

Firstly, for feasibility, the study focused on data governance frameworks created within the last ten years (2013-2022), potentially missing out on valuable insights from earlier frameworks that could inform current practices. However, the research team considers that analyzing recent frameworks would offer the most pertinent and current insights on data governance practices. While the exclusion of older frameworks may omit some valuable insights, this limitation is deemed acceptable given the study's goal to provide current and relevant information.

Secondly, while the research aims to provide a comprehensive representation of data governance frameworks, the geographical scope is limited in terms of representation from the Global South. This may impact the generalizability of the findings, particularly for low- and middle-income countries.

Thirdly, the research prioritized public sector efforts over private-sector-oriented ones, cognizant of how the public sector plays a critical role in setting policies and regulations for data governance. However, the importance of the private sector in this area cannot be overlooked, and further research may be necessary to fully understand data governance in the private sector context.

Finally, the research was initially conducted to inform the UN HLCP Group, which is primarily focused on issues related to the UN's programs and strategies for sustainable development. While the sample includes a wide range of frameworks and approaches, encompassing principles, guidelines, and regulatory frameworks and standards, the prioritization of frameworks and policies relevant to humanitarian and development issues may limit the applicability of the findings to other sectors.

In conclusion, the study remains relevant to a wide range of stakeholders interested in data governance. Indeed, while the study's focus and limited scope may restrict the generalizability of the findings, the research team implemented a systematic and rigorous approach to provide valuable insights into contemporary practices and challenges in data governance. As such, the study's outcomes can inform and guide future research and policy-making endeavors in this critical field, demonstrating its significance and usefulness to various audiences.

# 4. Main findings

The following section summarizes the key findings of the research with respect to existing data governance frameworks leveraging the previously-described analytical framework.

## 4.1. Purpose

**Key Takeaways:**

- Every reviewed data governance framework explicitly mentions its purpose.
- Goals and objectives as it relates to data governance were either sector specific or expressed general, broader aims.
- An emerging purpose this research identified across all frameworks, is the reconciliation of the tension between data protection and the increased use of data for societal goals.

All of the reviewed data governance frameworks include an explanation of the overall purpose of the document. From the sample, there is great variety in the scope of the purposes identified by the different frameworks. This research identified two types of purposes: those that refer to specific cases and sectors where data is used, and those that aim to improve data governance in general. Often, the latter are pursued by national or local governments seeking to achieve responsible use and reuse of data.

As for the former, for instance, the International Committee of the Red Cross (ICRC)'s "Handbook on Data Protection in Humanitarian Action" seeks to raise awareness and assist humanitarian organizations in ensuring that they comply with existing personal data protection standards in carrying out humanitarian activities specifically. As for the latter, on the other hand, the Personal Information Charter developed by the Foreign, Commonwealth & Development Office (FCDO) in the United Kingdom has very data-focused objectives that don't refer to one specific field (e.g. the humanitarian sector). Indeed, the charter provides the standards people can expect from the FCDO when they ask for, or hold, people's personal information.

Overall, an overarching purpose this research identified across all frameworks, regardless of their scope, is to balance the tension between the importance of *protection* against unauthorized collection and potential misuse of data versus the wider *promotion* of data for advancing various public interest goals. In fact, the frameworks identified often stemmed from the dual realization that data is indeed very valuable and potentially beneficial for public purposes, on the one hand, and that at the same time it poses a series of challenges and risks across its life cycle. Indeed, the promotion of data has the potential of bringing about a series of public benefits, spanning from more efficient mobility (Lau 2020) to personalized healthcare (Morgan 2021), from improved waste management (Abdallah et al. 2020) to more accessible education (Marchant 2021). However, the misuse of data can result in potential issues including social exclusions (O'Neil 2016) and injustices (Couldry and Mejias 2019), wasted time and resources (Redman 2016), as well as privacy (Cohen 2012) and legal concerns (Rodrigues 2020).

As a result, many governance frameworks seek to develop a variety of approaches aimed to leverage the new opportunities data presents, while avoiding the risks of its misuse. This appears to be a balancing act that is necessary but difficult to accomplish. Indeed, these efforts can result as being fragmented and difficult to operationalize.

One way to tackle such fragmentation is through data stewardship, which The GovLab defines as "policies, functions and competencies to enable access to and re-use of data for public benefit in a systematic, sustainable, and responsible way" (Verhulst 2021a). Similarly, the Ada Lovelace Institute defines it as "responsible use, collection and management of data in a participatory and rights-preserving way" (Ada Lovelace Institute 2021). The concept of stewardship has been used by Nobel Prize-winning economist Elinor Ostrom to describe the governance of common resources (Ostrom 2012). When considering data as a common good, Ostrom's design principles and conceptualization of stewardship can be a useful tool to find a balance between data protection and data promotion. Data stewards can play a crucial role in steering the process of using data and the insights it can generate by deciding who has access to data, for what purposes and to whose benefit (Open Data Institute 2022), ultimately addressing society's biggest questions and challenges in a systematic and responsible way (The GovLab 2018).

## 4.2. Principles

**Key Takeaways:**

- The research identified three main themes with respect to principles: (a) trust, (b) individual rights and interests, and (c) public interest .
- Among the principles used, Confidentiality and Security, Accountability, and Transparency were the three most mentioned principles, and all of them build on the earlier Fair Information Practice Principles (FIPPS).
- There is a lack of clarity and harmonization of meanings, especially across different countries. This makes it difficult to operationalize the principles.
- Three main types of principles were identified, namely principles for processes, principles for decisions, and principles for handling data.

### 4.2.1. Main Themes

In examining the principles across the governance frameworks in the sample, this research identified three main themes:

- **Trust:** Trust is essential both as a right in and of itself and moreover to enable the adoption and widespread usage of technologies and data platforms. Accordingly, a large number of the frameworks under examination emphasize trust as a central principle, both on ethical and practical grounds. An example can be found in the Privacy Impact Assessment developed by the Office of the Australian Information Commissioner, which emphasized a value proposition to strengthen community trust in the data initiatives carried out by the government. Another example, at the international level, are the WHO Data Principles. These highlight that their aim is to "provide a foundation for continually reaffirming trust in WHO's information and evidence on public health" (World Health Organization (WHO) 2020, 1).

- **Individual rights and interests:** Protecting citizen and user privacy rights emerges as one of the central principles in the frameworks analyzed in the sample, spanning across sectors and geographies. For example, the International Organization for Migration's Data Protection Manual emphasizes a core value proposition to "assist IOM staff to take reasonable and necessary precautions in order to preserve the confidentiality of personal data and ensure that the rights and interests of IOM beneficiaries are adequately protected" (International Organization for Migration 2015, 3).

- **Public interest:** A large number of frameworks under examination also emphasize the importance of increasing the scope of use of data so that it can be deployed more widely, in service of various public interests. For instance, the Data Sharing Policy of Médecins Sans Frontières (MSF) emphasizes that the organization's repository of data "can potentially be of value to researchers working in public health" (Médecins Sans Frontières (MSF) 2013, 4).

### 4.2.2. Most Used Principles

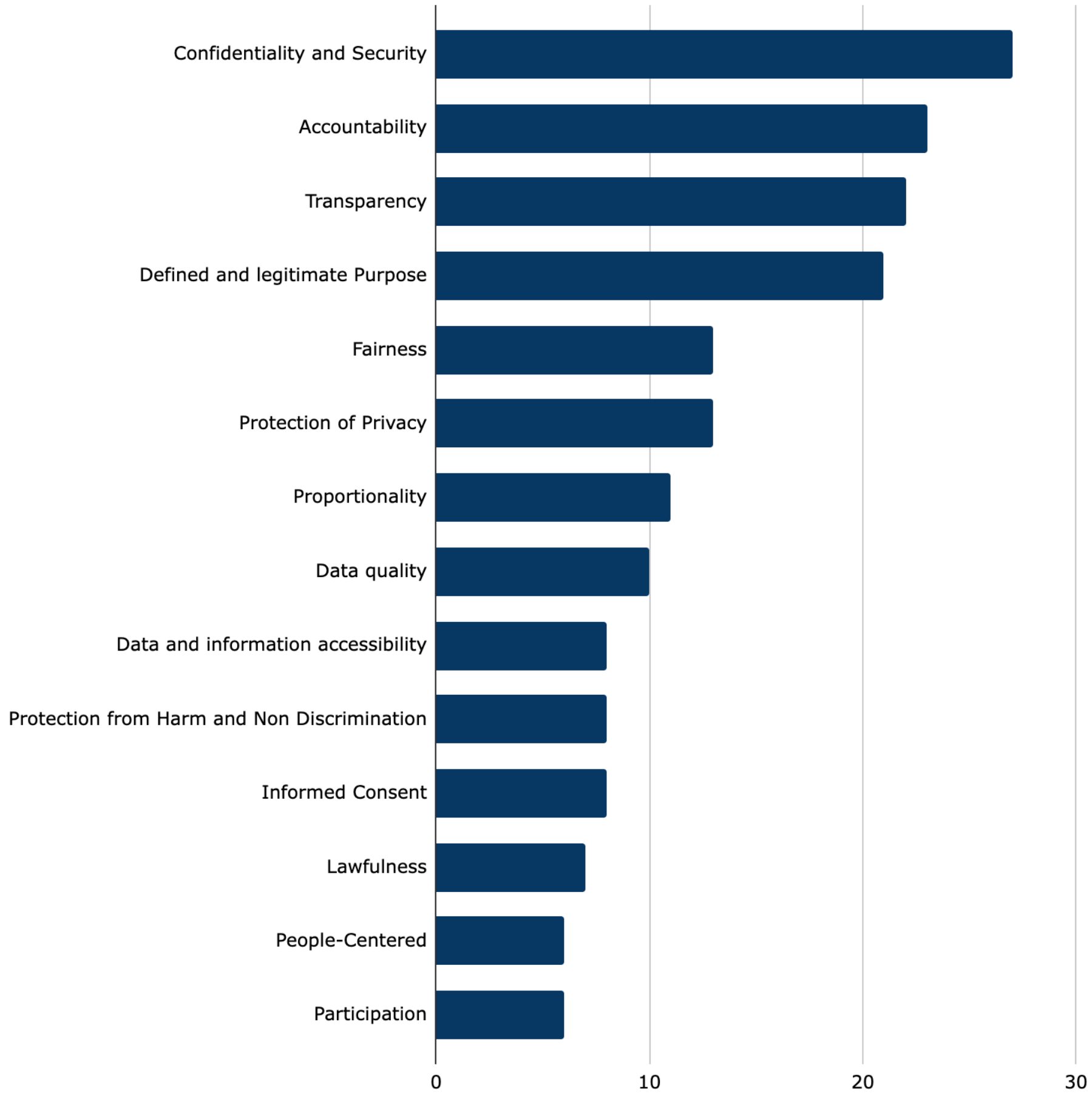

*Figure 8: Most used principles.*

This research identified the following findings with regards to the frameworks examined in the sample:

- **Lack of clarity and harmonization of meanings**
  - A large number of the frameworks include similar principles, but use different nomenclature (thus increasing the challenges of cross-framework comparison and analysis). For instance, the principle of "equity" is absent in many frameworks, but it seems to be implied under the principles of "fairness" or "non-discrimination".

- Moreover, a large number of frameworks do not separate principles, but group them together. Examples of principles often mentioned together are “necessity and proportionality” and “legitimate and fair processing”.

- **Different meanings in different countries**
    - The difference in nomenclatures –and substantive definitions of similar concepts– is especially noteworthy across countries and geographic regions. For instance, “privacy” is defined and applied differently in different countries. How can that difference be reconciled under a global data governance framework?

- **Association with Fair Information Practice Principles**
    - Many of the most used principles are associated with the Fair Information Practice Principles. Among those, this research found: Confidentiality and Security, Purpose specification, Transparency, Accountability, Data & Information Accessibility, Data Quality, Proportionality, Participation.

### 4.2.3. Processes, Governance, and Handling Principles

This research identified different categories of principles that sought to inform and steer different aspects of the data governance life cycle i.e. processes, decisions and data handling (albeit there is overlap):

- **Principles for Processes**, which include principles whose aim is to shape the processes followed to arrive at certain governance decisions. These include:
    - Transparency
    - Accountability
    - People-Centered
    - Fairness
    - Participation
    - Lawfulness

- **Principles for Decisions**, which include principles whose aim is to shape the governance decisions themselves. These include:
    - Transparency
    - Proportionality
    - Defined Purpose
    - Accountability
    - People-Centered
    - Fairness
    - Protection from Harm

- and Non-Discrimination
        - Participation

    - **Principles for Data Handling**, whose aim is to influence the way data is processed and handled. These include:
        - Confidentiality and Security
        - Proportionality
        - Data and Information Accessibility
        - Protection of Privacy
        - Lawfulness
        - Informed Consent
        - Data Quality

## 4.3. Anchoring

**Key Takeaways:**

- 63% of the frameworks analyzed mentioned anchoring documents that they built on and referred to as a point of reference.
- Anchor documents are mainly referred to as starting points, instead of binding documents to comply with.
- Of all the approaches analyzed, only 39% explicitly mentioned universal human rights frameworks.
- Two specific types of anchoring documents were identified in the sample: (a) international data and/or human rights protection standards, (b) previously-established privacy legislation.
- A minority of the frameworks considered did not specifically refer to any legal basis.

Overall, 63% of the frameworks analyzed mention anchoring documents that they built on and referred to as a point of reference. Of those, 59% built on international human rights norms and principles. This means that only 39% of the analyzed approaches explicitly mention universal human rights frameworks. This research observed that anchor documents are mainly referred to as starting points instead of binding documents to comply with. In this sense, none of the analyzed frameworks mention roles or groups responsible for overseeing compliance with said anchor document. However, there is a relevant caveat in this context since many actors that authored the analyzed frameworks are international and supranational organizations that might possess specific privileges and autonomy to act and decide (Reinisch 2009) and, therefore, have freedom in compliance with potential anchor documents.

Overall, the analysis identified three main types of frameworks in relation to anchor documents:

- **Frameworks that refer to international data and/or human rights protection standards**
    - An example of a framework that refers to international data protection standards is the "Data Strategy of the UN Secretary-General for Action by Everyone, Everywhere with Insight, Impact and Integrity", which emphasizes respect for human rights as well as international standards, such as the "UN Personal Data Protection and Privacy Principles" (United Nations 2020, 60). Another example is the "Signal Program on Human Security and Technology" at the Harvard Humanitarian Initiative, which builds on rights identified in the Universal Declaration of Human Rights (UDHR), the International Covenants on Civil and Political Rights (ICCPR), and other instruments of humanitarian law rights that apply to all people "regardless of the use of any specific technology" (Harvard Humanitarian Initiative 2015, 8).

- **Frameworks that refer to previously-established privacy legislation**
    - An example of a framework that builds on pre-existing privacy regulation is the Australian Information Commissioner's Guide for Privacy Impact Assessment, which evaluates compliance to previously-established privacy legislation such as the Privacy Act 1988. Another example is the United Kingdom's Personal Information Charter, which seeks to ensure that all personal information is treated in accordance with the UK General Data Protection Regulation and the Data Protection Act from 2018.

- **Frameworks that don't refer to any specific legal basis**
    - Finally, a minority of the frameworks considered did not specifically refer to any legal basis. These were mainly collections of principles and general data governance guidelines, such as the "Data Privacy, Ethics and Protection: Guidance Note on Big Data for Achievement of the 2030 Agenda" developed by the UN Development Group (UNDG). This document sets out general guidance on data privacy, data protection and data ethics for the United Nations Development Group (UNDG) concerning the use of big data collected in real time by private sector entities, and was shared with UNDG members for the purposes of strengthening operational implementation of their programmes to support the achievement of the 2030 Agenda. The guidelines do not refer to any legal basis, and provide a minimum base for self regulation.

## 4.4. Data Description

**Key Takeaways:**

- 51% of the frameworks analyzed clearly define the type of data they aim to oversee.

- 49% of the frameworks have an unclear or only partial definition. These often define data in general but don't provide a definition of the data they oversee.
- Emerging concepts such as sensitive and synthetic data are often missing, as well as concepts associated with relational data, group privacy, and collective rights.

This research analyzed how data was defined and approached in the frameworks regarding the type of data (data description) and the data lifecycle. Concerning data type, an examination of the sample results in the following observations:

- 51% of the frameworks analyzed clearly define the type of data they aim to oversee. Nearly all of the frameworks that clearly define data are involved with the oversight of personal data. Indeed, 90% of those frameworks oversee personal identifiable data. Although there is a lack of consensus about a unique data definition, some of the frameworks follow a definition close to the one proposed by the UN World Food Program: "Personal data is any information relating to an individual that identifies the individual or can be used to identify them" (UN World Food Programme (WFP) 2016, 2) which is similar in other UN agencies such as UNICEF, UNFPA, and International Organization For Migration.

- 49% of the frameworks have an unclear or only partial definition. These often define data in general but don't provide a definition of the data they oversee. For instance, the Responsible Data Management Training Pack by Oxfam follows the definition developed by the United Nations Economic and Social Commission for Asia and the Pacific (UNESCAP). Data is indeed defined as "the physical representation of information in a manner suitable for communication, interpretation, or processing by human beings or by automatic means. Data may be numerical, descriptive or visual" (Oxfam 2015, 6).

- In general, the analysis found that emerging definitions and data types are missing from the governance frameworks. In particular, new concepts such as sensitive and synthetic data are often missing, as well as exploring relational data, which would lead to a broader discussion of collective or community rights.

## 4.5. Data Lifecycle

**Key Takeaways:**

- 24% of the frameworks acknowledge the relevance of the data lifecycle approach, develop it and provide specific recommendations for each stage while recognizing the different needs of each one.
- 38% of the frameworks only provide partial or general recommendations, often associated with limited scope or specific coverage of the data cycle.

- There is a noticeable lack of frameworks covering the data re-use stage.

When understanding data as a global asset, the concept of the "data lifecycle" is helpful for estimating the value of data. Value emerges in data transformation, from data collection, processing, and analysis into digital intelligence so that it can be monetized for commercial purposes or used for social objectives (UNCTAD 2021, 17). This process identifies the particular characteristics and requirements when managing data in each stage. The decisions at every stage of the data life cycle will vary, depending on the type of data and their proximity to features of public goods. Therefore, addressing data governance through a data lifecycle approach is helpful because it enables a comprehensive analysis of how data should be overseen at various stages and create value from data use and reuse in a safe and equitable manner. Figure 9 presents a graphic representation of the data lifecycle.

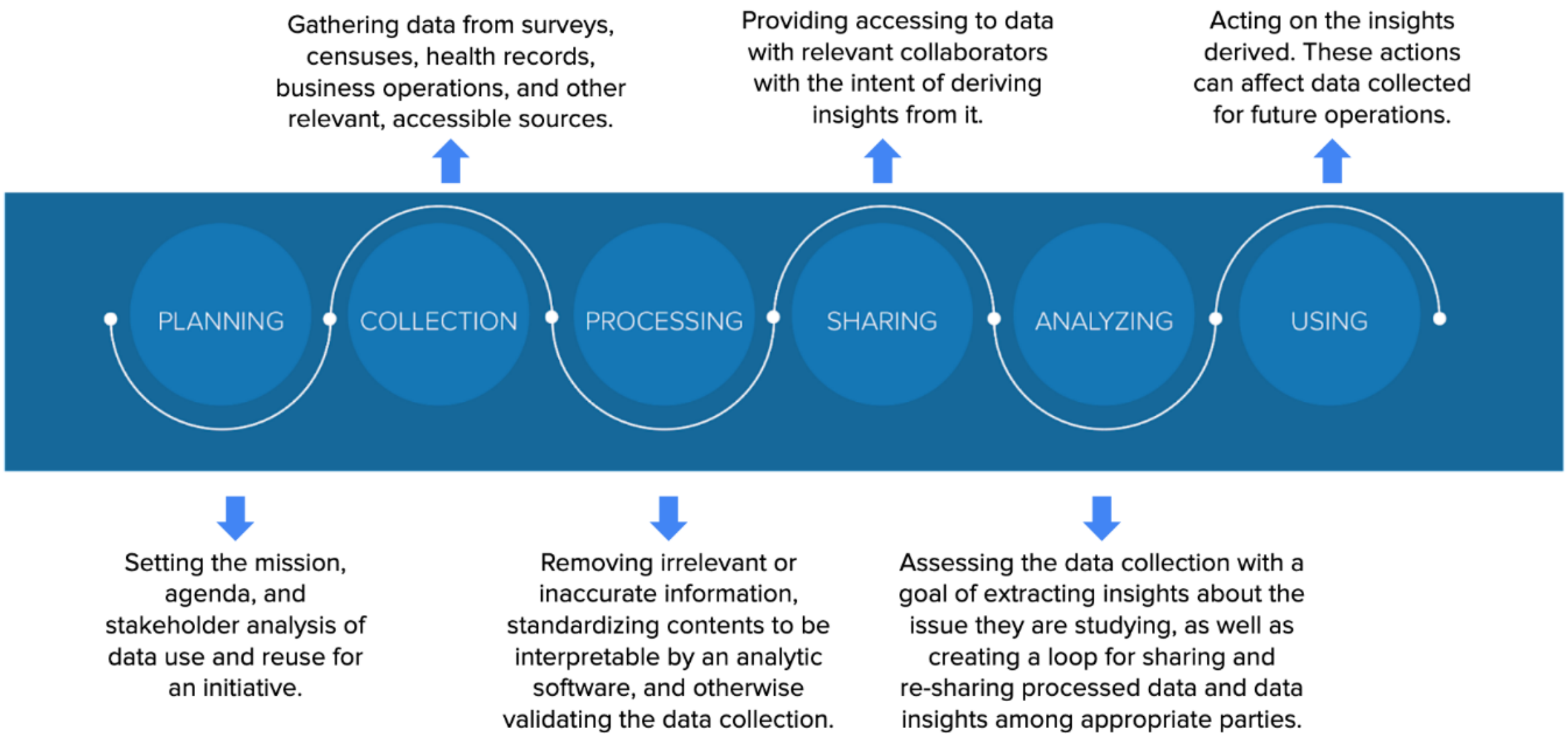

*Figure 9: Data Lifecycle by The GovLab.*

When analyzing the sample frameworks from a lifecycle approach, the research found:

- Only 24% of the frameworks acknowledge the relevance of the data lifecycle approach, develop it and provide specific recommendations for each stage while recognizing the different needs of each one.
    - A good example of those is the "ASEAN Data Management Framework". It has been developed to support one of the four strategic priorities identified in the ASEAN Framework on Digital Data Governance, the Data Life Cycle & Ecosystem, to define data governance throughout the data lifecycle (e.g., collection, use, access, storage). The framework then provides adequate data protection

recommendations for different data types within an organization and throughout the data lifecycle.
    - Another example from a sectoral lens is the International Civil Aviation Organization (ICAO)'s "Doc 8126 Aeronautical Information Services Manual". It recognizes the different stages of data under the Aeronautical Information Management concept that compiles acquiring aeronautical data, processing (validation, verification, and management) aeronautical data and information, providing access to aeronautical information through information services, and consuming aeronautical information by the end users. For each of those stages, the framework gives specific recommendations.
    - For instance, the Responsible Data for Children Synthesis conceptual framework from UNICEF and The GovLab describes the data lifecycle in six broad stages: planning, collecting, storing and preparing, sharing, analyzing and using, and provides actionable insights to work with children's data in the advancement of responsible practices through the data lifecycle.

- 38% of the frameworks mention (directly or implicitly) the data lifecycle approach, yet provide partial or general recommendations. In some cases, it is associated with limited coverage of the data cycle. In general, existing governance frameworks tend to focus mainly on using data for the purpose collected (planning stage). In contrast, only four analyzed frameworks cover data localization requirements (processing stage). There is a noticeable dearth of frameworks covering the data re-use stage.
    - For instance, the Data Security Law of the People's Republic of China acknowledges the data lifecycle by stating that "Data handling" includes the collection, storage, use, processing, transmission, provision, and disclosure of data. However, the law does not provide particular recommendations or guidelines for each stage.
    - In the case of the GDPR, for example, there is no explicit differentiation between the stages of data processing since "processing" is defined as any operation performed on personal data "*such as collection, recording, organization, structuring, storage, adaptation or alteration, retrieval, consultation, use, disclosure by transmission, dissemination or otherwise making available, alignment or combination, restriction, erasure or destruction"* (*General Data Protection Regulation* 2016, Article 4(2)) . Although it covers different data lifecycle phases under the processing definition, it does not provide specific recommendations based on the data cycle stages. This approach applies throughout the regulatory framework, except for the chapter on "transfers of personal data to third countries or international organizations" (*General Data Protection Regulation* 2016, Chapter V), where the focus is data sharing.
    - For example, among the frameworks with limited scope, the Data Sharing Policy of the Médecins Sans Frontières (a non-governmental organization) stands out. This policy aims to guide the use of health data generated by its programs. Even

though it recognizes the different stages of the data lifecycle, the only concern of this policy is the data-sharing stage.

- Finally, 38% of the frameworks do not identify nor acknowledge explicitly the data lifecycle and its different stages to develop their approach to data governance.

### 4.6. Processes

**Key Takeaways:**

- Only 5% of the frameworks followed a participatory process that included data subjects when defining the data governance approach.
- 22.4%of the frameworks created a specific governance body with functions to oversee framework implementation. The rest added extra functions to existing agencies or authorities or made vague high-level recommendations.
- Only 29% of the sample explicitly state how supervisory authorities will monitor and evaluate compliance with the framework. 29% vaguely recommend establishing a monitoring mechanism, and 41% did not establish or mention the need for monitoring mechanisms.

A well-functioning data governance framework should define the roles, responsibilities, and associated compliance procedures to safely share, use, and reuse data by all stakeholders. This research identified three types of processes:

- **Process to develop the governance framework:** Move toward more participatory processes.
    - In at least 18% of the sample, the analysis found that multi-stakeholder and multisectoral approaches were used to define frameworks, especially those within intergovernmental institutions and international coalitions. For example, the OECD's "Recommendation of the Council on Health Data Governance" results from a multi-stakeholder effort. It was jointly developed by the Committee on Digital Economy Policy and the Health Committee, the former Working Party on Security and Privacy in the Digital Economy (renamed in 2019 as the Working Party on Data Governance and Privacy), and the former Health Care Quality Indicators Expert Group (OECD 2016). However, it did not involve any data subjects in the process.
    - Only 5% of the frameworks followed a participatory process that included data subjects when defining the data governance approach. For instance, and perhaps the most representative example of participation, is the Data Governance Framework for New Zealand, where the government co-designed the framework with the Māori community (data subjects) to reflect Māori needs and interests in data. Another example is UNESCO's Internet Universality Indicators which were

developed in three phases, including two rounds of consultations, consultative meetings and workshops at international, regional, and national events with a diverse group of stakeholders, including civil society and individuals.
    - In general, the remaining frameworks displayed shortcomings with regard to inclusiveness.

- **Process to identify and create new professions and functions:** Need for more specific roles and binding responsibilities.

    - Roughly a fourth (22.4%) of the frameworks created a specific governance body with functions to oversee framework implementation. Titles for these bodies varied, and included such designations as Data Protection Officers and Chief Data Officers.
        - For example, the "Guidance on the Protection of Personal Data of Persons of Concern to UNHCR" creates new roles and responsibilities: the Data Controller, the Data Protection Officer and the Inspector General's Office, and the Ethics Office with specific responsibilities to oversee the compliance of the policy. The data controller is responsible to establish procedures that respect the rights of data subjects. The data protection officer supervises and monitors the compliance with the Data Protection Policy. The Inspector General's Office must investigate complaints of data subjects under the right to information. And, the Ethics committee must provide a whistle blower policy with respect to data protection (UN High Commissioner for Refugees (UNHCR) 2018).
        - In the case of the GDPR, for example, each member state must establish an independent public authority responsible for monitoring the application of the GDPR, which is called "supervisory authority". Further, a lead supervisory authority is established for cross-border data processing. The supervisory authorities are designed to monitor and enforce the regulation on their territory. Moreover, the GDPR establishes a European Data Protection Board, composed of the heads of the supervisory authorities of each member state. In addition, each data controller or processor must designate a data protection officer.
        - The California Consumer Privacy Act (CCPA), for instance, created the California Privacy Protection Agency, which is vested with full administrative power, authority, and jurisdiction to implement and enforce the CCPA. It also specifies that the agency will have a five-member board, including a chairperson. "The chairperson and one member of the board shall be appointed by the Governor. The Attorney General, Senate Rules Committee, and Speaker of the Assembly shall each appoint one member. These appointments should be made from among Californians with

expertise in the areas of privacy, technology, and consumer rights" (California Consumer Privacy Act 2018, Section 1798.199.10) .

- In contrast, 24.1% of the sample simply added extra data governance functions to existing agencies or authorities, often without the required capacity- and expertise-building.
  - The Office for the Coordination of Humanitarian Affairs (UN OCHA) data responsibility guidelines, for example, establish that OCHA Centre for Humanitarian Data, an existing body, "is committed to supporting offices and sections across OCHA in adopting the Guidelines" (OCHA Centre for Humanitarian Data 2021, 34).
  - For instance, the "Responsible Program Data Policy" by Oxfam (non-governmental organization) explicitly states that overseeing the framework implementation must be executed by the Oxfam Country Directors as an additional task.
  - Also, the "Procedures for Ethical Standards in Research, Evaluation, Data Collection and Analysis" by UNICEF present a similar approach. It states that Country Representatives, Regional Directors, and Heads of Divisions must ensure and maintain "the highest ethical standards in all evidence generation endeavors" (UNICEF Division of Data, Research and Policy 2015, 2) by implementing the procedures laid out in the framework.

- A small number of other frameworks (19%) recommend the establishment of roles and governance bodies, but do not actually include provisions for creating them as part of non-binding recommendations.
  - The OECD's Recommendation of the Council on Health Data Governance provides vague instructions to existing governance bodies given its supranational nature. It recommends governments engage with relevant experts and organizations to develop mechanisms to implement the framework. It also encourages non-governmental organizations to follow the recommendation when processing personal health data for health-related purposes that serve the public interest. However, the recommendations are not legally binding for member states or non-governmental institutions.
  - For example, the Association of Southeast Asian Nations (ASEAN) Data Management Framework (DMF) advises members to define roles and responsibilities for implementing each action described in the framework: "To develop and implement the 6 foundational components of the DMF, an organization is required to identify and determine different roles and responsibilities in order to ensure adoption, operation, and compliance, in accordance with business needs" (Association of Southeast Asian Nations (ASEAN) 2021, 12). However, the framework is not specific regarding the new professional roles and skills required.

- There is another group of frameworks that do not provide governance information nor propose new roles or tasks (34.5%), mainly because of the nature and scope of the document.
  - This usually happens when looking at principles frameworks such as the Gemini Principles, the CARE Principles for Indigenous Data Governance, or the UN High-Level Committee on Management Personal Data Protection and Privacy Principles. None of those specify or suggest establishing governance roles since they mainly focus on the values when managing data. Or in the case of the Risks, Harms and Benefits Assessment by UN Global Pulse, for example, given that it is a data privacy and protection compliance mechanism, the scope limits its role to provide governance recommendations.

- **Processes to monitor and evaluate:** Highly related to governance roles and functions, these elements that fell short are generally insufficiently defined.
  - 41.4% of the frameworks did not establish or mention the need for monitoring mechanisms, which are a vital component of successful operationalization for any framework.
    - For example, the “Handbook on Data Protection in Humanitarian Action” created by the Brussels Privacy Hub and the International Committee of the Red Cross (ICRC) recognizes that international organizations have complete independence on how to process data and monitor compliance of data protection recommendations “they can therefore process Personal Data according to their own rules, subject to the internal monitoring and enforcement of their own compliance systems; in this regard they constitute their own jurisdiction" (International Committee of the Red Cross, 35) .
    - On the other hand, the considerations for Using Data Responsibly for the United States Agency for International Development (USAID) acknowledge that this is not an enforceable framework and define it as an internal document to start the conversation on data governance.

  - Some of the frameworks (29.3%) recommend establishing a monitoring mechanism, although it is often unclear whose responsibility it is to ensure those mechanisms are implemented.
    - For instance, the International Organization for Migration's (IOM) Data Protection Manual acknowledges the importance of oversight and compliance. Still, it only advises creating, without appointing, "an independent body to oversee the implementation of these principles and

to investigate any complaints, and designated data protection focal points should assist with monitoring and training" (International Organization for Migration 2015, 12).

    - The ICAO's Aeronautical Information Services Manual, for example, recommends establishing monitoring and evaluation mechanisms but falls short of providing real instruments or mandates. "States must implement well-documented surveillance processes by defining and planning inspections, audits, and monitoring activities on a continuous basis" (International Civil Aviation Organization 2021, Section 2.8.1).

- The other 29.3% of the sample explicitly state how supervisory authorities are going to supervise and monitor the compliance of it. Some of them might correspond to regulatory frameworks or laws.
    - Canada's Personal Information Protection and Electronic Documents Act (PIPEDA), for instance, establishes that the Office of the Privacy Commissioner of Canada will oversee compliance with the PIPEDA legislation. Further, the Commissioner may audit an organization's personal information management practices if the Commissioner believes that the organization has not followed a recommendation set out in the Act; it may ask for additional resources to monitor implementation.
    - Similarly, the GDPR establishes in Articles 41 and 42 that a supervisory authority for monitoring compliance may be carried out by a body with an appropriate level of expertise in relation to the subject matter of the code and is accredited for that purpose. That way, the Member States, the supervisory authorities, the Board, and the Commission shall encourage compliance with this Regulation of processing operations by data controllers and processors (General Data Protection Regulation 2016).
    - Another example is the Recommendation of the OECD council on health data governance. In this case, the framework provides detailed guidelines to implement monitoring and evaluation mechanisms, such as assessing whether the uses of personal health data have met the intended health-related public interest purposes and brought the benefits expected by i) pursuing a periodic review of developments in personal health data availability, the needs of health research and related activities, and public policy needs; and i) following a systematic assessment and updating of policies and practices to manage privacy, protection of personal health data and security risks relating to personal health data governance.

- In addition, none of the frameworks under study established evaluation mechanisms, which makes it difficult to assess their impact or effectiveness.

## 4.7. Practices

**Key Takeaways:**

- Only 29% of frameworks provided practical tools such as templates, checklists, or assessments.
- 38% of the frameworks analyzed recommend and describe, often in a detailed fashion, good practices for data governance. But they fail to define who, how, and when these practices should be implemented.
- With few exceptions (e.g., data protection regulatory frameworks and laws), most recommended practices are not binding. This may limit their effectiveness and the extent to which they are operationalized.

This benchmarking exercise examined the sample frameworks for their practices, which aimed to analyze their practical approach, and how they translate the theory, values, and principles into practice. From applied templates and checklists to the description of processes, sometimes imprecise, four trends were observed:

- **Plethora of practices but often vague**
  - 38% of the frameworks analyzed recommend and describe, often in a detailed fashion, good practices for data governance. However, in most cases the frameworks do not include detailed recommendations or guidelines for who, how, and when these practices should be implemented.
    - The California Consumer Privacy Act (CCPA), for example, provides guidance to businesses on how to inform consumers of their rights under the CCPA, how to handle consumer requests, how to verify the identity of consumers making requests, and how to apply the law as it relates to minors. It stipulates the processes businesses need to implement to follow the CCPA while making it easier for consumers to exercise their CCPA rights (California Consumer Privacy Act 2018).
    - Another example is the UNICEF & UNFPA Policy on Personal Data Protection. The Policy describes good practice when collecting data from individuals and explains how to notify the data subject. It presents the information that shall be provided to each identified data subject within a reasonable period when personal data are collected by UNICEF or UNFPA (as controller), taking into account the logistical constraints both organizations face (UNICEF 2020).
    - For example, the Privacy Impact Assessment Toolkit from the Office of the Australian Information Commissioner describes a detailed ten-step process for undertaking a privacy impact assessment to apply the toolkit. Similarly, the Data responsibility guidelines of the UN Office for the Coordination of Humanitarian Affairs (UN OCHA) recommend eight actions

to be implemented at the system-wide, sector, and organizational levels. The framework guides OCHA Staff to implement these processes at different levels. However there is no clarity regarding the timing and the ways in which the actions should be implemented.

- **Need for practical tools such as templates for sharing agreements**
  - Data collection and data sharing are key components of the data life cycle, but rife with potential minefields (e.g., potential regulatory or privacy violations). Organizations, especially if they are under-resourced technically or financially, can greatly benefit from detailed guidelines such as templates and checklists to help guide their decisions. In the sample, only 29% of frameworks provided tools such as templates, checklists, or assessments.
    - One of them is the International Organization for Migration's Data Protection Manual that provides a number of practical tools: templates and checklists. Templates of model consent forms, general contractual clauses to be inserted into contracts, and request forms for data subjects seeking access to their personal data; and checklists for data quality, data security, and data protection.
    - Also, the USAID's Considerations for Using Data Responsibly offers applied tools for how to help guide discussions or navigate areas of responsible data practice that may be unclear. The tools range from the key events planning table to the benefits risk assessment, a worksheet to track and protect copies of sensitive data and IT security highlights checklist.
    - The UNDP's Data Principles, even though they are principles, provide practical resources to implement their recommendations. For example, the Informed Consent explanation for Safeguarding personal data, the Responsible Development Data Book for Manage Data Responsibly, and the Mozilla Science Data Reuse Checklist for planning for reusability and interoperability.
    - For example, the "WFP Guide to Personal Data Protection and Privacy" offers a Self-Assessment Compliance Checklist that allows personnel to measure compliance with each of the elements of the guidelines. Further, it presents a Model Consent Forms that can be used to develop local templates for obtaining informed consent and responding to beneficiaries' requests for access to their data.

- **Risk assessments and compliance mechanisms**
  Within the processes and tools presented in the analyzed frameworks, there is a subgroup of risk assessments that stand out as a good practice for applying responsible data management.

- For example, the Risks, Harms and Benefits Assessment of the UN Global Pulse develops a two-steps assessment: the first one is a checklist, which is used as an initial assessment tool that identifies potential risks and helps evaluate whether a more comprehensive review should be conducted; and the second one consists of a detailed measurement of the likelihood, magnitude, and significance of impacts of a data innovation project if a medium or high risk was identified in the initial assessment.
    - Also, the “Handbook on Data Protection in Humanitarian Action” by the Brussels Privacy Hub (VUB) and ICRC offers a Data Protection Impact Assessment (DPIA) tool to identify, evaluate and address the risks to Personal Data arising from a project, policy, program, or another initiative. It includes a step-by-step guide for humanitarian organizations to conduct it. It also has a template for a DPIA report.
    - The “Considerations for Using Data Responsibly at USAID”, among the other provided tools, develops the benefits risk assessment, a tool designed to help assess potential benefits and risks of data collection, use, and sharing. To properly evaluate risks and benefits, it is critical to include relevant stakeholders in this process, including those from whom you collect data.

- **Non-binding frameworks**
    - With few exceptions (e.g., data protection regulatory frameworks and laws), most recommendations are not binding. This may limit their effectiveness and the extent to which they are operationalized. Their list of recommendations and good practices are “suggestions” or “proposals.”
    - Furthermore, 34% of the data frameworks reviewed do not provide detailed processes or explicit practical tools.

# 5. Final Recommendations

Data governance embraces a wide range of elements and concepts without a unified or unique definition, which might create asymmetries when establishing a data governance framework. The analysis has shown a variety of purposes, approaches and scopes at the local, national, and international levels. Davis proposes a definition aiming to gather multiple components of it: “Data governance concerns the rules, processes and behaviors related to the collection, management, analysis, use, sharing and disposal of data – personal and/or non-personal. Good data governance should promote benefits and minimize harms at each stage of relevant data cycle” (Davis 2022, 12).

Well-designed data governance, according to the World Bank , can be defined as the framework that allows capturing the central values and purposes of an entity (country, international body, region, etc.) to leverage the synergies with multiple stakeholders while creating trust and promoting the use of data (World Bank 2021, 10). Based on those definitions and building upon the main takeaways from the detailed analysis of 58 data governance frameworks, there is an opportunity for researchers, decision-makers and other stakeholders to identify critical elements and follow good practices. The following reflections are clustered according to the proposed analytical framework (Section 3.3.):

1. **Consider data stewardship to reconcile the tension between data protection and data promotion**: Moving forward, there may be a need to adopt a broader framework and concept of data stewardship. This would indeed allow to achieve and maintain the dual goal of protecting and promoting data in a more systematic, sustainable and responsible way. Indeed, as mentioned in Section 4.1., data stewardship aims to make the use of data more responsible, systematic and sustainable (Verhulst 2021a); achieve the responsible and accountable use of common resources (Ada Lovelace Institute 2021) allowing to make full use of data's benefits and avoiding the social and economic harms that can stem from its misuse (Open Data Institute 2022). In particular, it may be useful to (a) Provide a legal, shared definition for global data stewardship, (b) Rationalize and coordinate existing support to international data stewardship efforts, and (c) Commission research and trials to assess the potential of a global data stewards association.

2. **Focus on responsible re-use to unlock the socioeconomic value of data:** In recent years, the open data movement to improve public governance has grown significantly (The GovLab 2016). As a consequence, increasing amounts of both public and private data have been made available to external stakeholders. However, although the frameworks analyzed in this research did aim to develop different ways to govern the use of data, they overall lacked a focus on the re-use of data–i.e., the sharing of data across different domains. It may in fact be beneficial to integrate the concept of reuse in the development of a global data governance framework, so as to create shared approaches and standards with respect to the sharing of data amongst different stakeholders. In particular, it may be useful to (a) Develop methodologies to define and measure the value of data, (b) Develop structures to incentivize the 'co-creation of value' (Mazzucato 2019), (c) Encourage data collaboratives,[3] and (d) Identify and nurture data stewards, as further specified in Reflection 1 (Verhulst 2020). Finally, in order for the reuse of data to be deemed responsible and consequently legitimate, it is crucial to create avenues for public assemblies and value the importance of social licenses (see Recommendation 7) (Verhulst 2021b).

3. **Harmonize meanings to operationalize principles:** This research showed that there is an overall lack of clarity and harmonization of meanings across different countries, sectors, and organizations. This makes it difficult to operationalize the principles in a harmonized manner and at a global level. Whereas different contexts are bound to value different principles to some degree, there seem to be an overarching agreement on a series of data governance principles (See Section 4.2.2.). These, however, seem to be defined differently by different organizations. It may be worth universalizing the principles to be embedded in a global data governance framework, so as to systematically operationalize them and ensure compliance across different regions and nations.

[3] In this research, 'data collaboratives' are defined as "...an emerging form of public-private partnership that enables sharing and co-creation of value. They may involve, for instance, informal and time-bound collaborations between a company and an academic research group or civil society organization, and allow data to be re-purposed, typically in an anonymized form and with specific intent." (Verhulst 2020).

4. **Use broader anchoring frameworks to provide common North Stars:** As mentioned in Section 4.2., of all the approaches analyzed, only 39% explicitly mention global human rights frameworks. Moreover, anchor documents are mainly starting points, instead of binding documents to comply with. Finally, because of the loose nature of the "anchoring process", none of the frameworks clearly mention responsible roles or identify processes for overseeing compliance with an anchor document. First, it seems that having broader frameworks–and not only, for instance, privacy-focused legal bases–related to universal human rights may be beneficial in developing a global data governance framework (MacFeely et al., 2022). Second, it seems important to establish clear levels of compliance required with such documents. Finally, based on those levels, it may be beneficial to answer the question of who oversees compliance, so as to ultimately materialize the relationship between the framework and its anchor document.

5. **Unify key definitions of data and incorporate emerging concepts such as synthetic data:** Whereas it is important to keep in mind that a fixed definition of data may be more harmful than beneficial, mainly due to the ever-changing nature of both data and the technologies it relates to, it is also crucial to develop a series of mechanisms that allow flexibility and clarity of what people and organizations mean by 'data'. In this sense, it may prove useful to incorporate emerging, flexible concepts such as synthetic data, as well as relational data, thick data, and sensitive data (with the latter having been increasingly more adopted lately). This indeed could enable the definitions to be more precise, without referring to the broad, general concept of data, and at the same time it may result in a malleable approach that could allow for the various data-related evolutions and developments to be assimilated.

6. **Adopt the data lifecycle approach to promote benefits and minimize harms:** Given the breadth of contexts in which data governance must be applied, it is beneficial to use a standardized framing to structure the needs, risks, and opportunities when handling data. As mentioned before, data governance results from multiple processes all aligned toward data promotion and protection. These processes can be hard to understand when viewed together, and although the data lifecycle is not linear, it could help to inform responsible data handling approaches better while promoting better and more impactful data management (The GovLab 2021).

7. **Incorporate more participatory processes and collective agency to develop a data governance framework:** Participatory data governance occurs when organizations allow different constituents to contribute to the discussion and are accountable for their decisions to the public. To encourage transparency and accountability in data governance efforts, decision-makers should offer opportunities for scrutiny and input from data subjects. This policy feedback process is particularly relevant within the data governance discussion since it will allow obtaining the most value from data while protecting people from harm (The Digital Trade and Data Governance Hub 2022). Therefore, public consultations on the design of policies and regulations could support transparency and stakeholder engagement (World Bank 2021, 284) while fostering the social license of the process. The social license refers to the informal permissions granted to institutions such as governments or corporations by members of the public to carry out a particular set of activities (Shaw, Sethi, and Cassel 2020), in this case, the collection, sharing, and use of their data.

   Most of today's participatory processes focus on protecting individuals' rights. Yet these debates fail to consider the agency of data subjects as a collective. There are often massive asymmetries between individuals and stronger stakeholders, such as the public or the private sector, that exploit their data while restricting its potential. To address these asymmetries, a new principle of digital self-determination is needed. Verhulst defines digital self-determination as "the principle of respecting, embedding, and enforcing people's and people's agency, rights, interests, preferences, and expectations throughout the digital data life cycle in a mutually beneficial manner for all parties involved" (Verhulst 2022, 10). In the context of data governance, building symmetric relationships can help data subjects leverage their self-determination more effectively to exert greater control over how their data is used and reused (Verhulst 2022). This is particularly relevant for underrepresented groups, which possess even lower bargaining power than other collectives.

8. **Invest in and create new professions with specific roles and responsibilities:** Attracting data talent and promoting data stewardship is key to ensuring compliance with data governance frameworks and fostering a culture around data collaboration and protection within organizations. The Third Wave of Data by The GovLab proposes a focus on new institutional arrangements to achieve a data-driven culture with particular attention on the role of the data steward - accountable data leaders that seek new ways to create value through cross-sector data collaboration (The GovLab 2021). The analysis reveals the importance of having trained and dedicated individuals (whether chief data, chief privacy, or chief security officers, or the equivalent body) with specific functions and binding responsibilities for long-lasting, sustainable, and informed data actions.

9. **Improve accountability and transparency by defining oversight and compliance mechanisms:** There is a need to explicitly define monitoring and evaluation mechanisms linked to defining roles and responsibilities. An organization's accountability can be measured by how it monitors and assesses its internal policies to manage, protect and secure data effectively (Centre for Information Policy and Leadership 2011) or simply by complying with the data governance policy. To do so, it is recommended to establish not only the mechanism but also who, when, how often, and how it should be implemented.

10. **Translate values and recommendations into practical tools:** In collaboration with diverse policy-makers and stakeholders, identify the most valuable tools to facilitate and accelerate the implementation of a data governance policy. The analysis identified three practical tools that should be considered to help data stewardship: model consent forms, checklists for data quality, data security and data protection, and data risks assessment step-by-step guidelines. These tools may indeed prove useful to guide implementation, document progress, and monitor compliance.

## Acknowledgements

This work would not have been possible without the support and guidance of many individuals throughout the research process. The authors would like to express our sincere gratitude to the peer reviewers for their valuable contributions and insights in reviewing this report. In particular, we would like to thank Dr Susan Aaronson, Research Professor at George Washington University and Director of the Digital Trade and Data Governance Hub, Jos Berens, Data Policy Officer at United Nations Office for the Coordination of Humanitarian Affairs (OCHA), and Malar Veerappan, Program Manager and Senior Data Scientist at the World Bank. Finally, we would like to thank the United Nations High-level Committee on Programmes (HLCP) Core Group on New Global Public Goods for their support and guidance throughout this research.

# Appendix

## Appendix A: List of Analyzed Frameworks

| Name of the Framework | Type | Owner/Author | Type of Organization | Geographical Scope | Sector | Year |
|---|---|---|---|---|---|---|
| Personal Information Charter | Charter | Foreign, Commonwealth & Development Office (FCDO) | Governmental Institutions | Local | Policy and Regulation | 2021 |
| Industry Toolkit: Children's Online Privacy and Freedom of Expression | Toolkit | UNICEF | Intergovernmental Institutions | Global | Humanitarian | 2015 |
| A Human-Rights Based Approach to Data: Leaving No One Behind in the 2030 Agenda for Sustainable Development | Conceptual Framework | UN Office for the Coordination of Humanitarian Affairs (UN OCHA) | Intergovernmental Institutions | Global | Humanitarian | 2018 |
| Data Strategy of the Secretary-General for Action by Everyone, Everywhere with Insight, Impact and Integrity | Conceptual Framework | United Nations | Intergovernmental Institutions | Global | Not One Specific Sector | 2022 |
| Principles for Digital Development | Principles | Principles for Digital Development | International Independent Coalitions | Global | Research and Development | 2017 |
| Responsible Data Program | Conceptual Framework | The Engine Room | International Independent Coalitions | Global | Research and Development | 2016 |
| Risks, Harms and Benefits Assessment | Conceptual Framework | UN Global Pulse | Intergovernmental Institutions | Global | Humanitarian | 2020 |
| DTM & Partners Toolkit: Enhancing Responsible Data Sharing | Toolkit | International Organization for Migration | Intergovernmental Institutions | Global | Migration | 2018 |
| Oxfam Responsible Program Data Policy | Toolkit | Oxfam | Non-Governmental Organizations | Global | Humanitarian | 2015 |
| Data Sharing Policy | Conceptual Framework | Médecins Sans Frontières | Non-Governmental Organizations | Global | Health | 2013 |
| Data Privacy, Ethics and Protection: Guidance Note on Big Data for Achievement of the 2030 Agenda | Manual/Guidelines | UN Development Group (UNDG) | Intergovernmental Institutions | Global | Research and Development | 2017 |
| Signal Program on Human Security and Technology at the Harvard Humanitarian Initiative. | Conceptual Framework | Harvard Humanitarian Initiative (HHI) | Academia and Research Institutions | Global | Humanitarian | 2015 |
| IOM Data Protection Manual | Manual/Guidelines | International Organization for Migration | Intergovernmental Institutions | Global | Migration | 2015 |
| Undertaking a Privacy Impact Assessment | Toolkit | Office of the Australian Information Commissioner | Governmental Institutions | Local | Policy and Regulation | 2020 |
| Responsible Data for Children Synthesis report | Conceptual Framework | UNICEF and The GovLab | Intergovernmental Institutions | Global | Humanitarian | 2019 |
| General Data Protection Regulation Framework | Regulation Framework | European Union | Intergovernmental Institutions | Regional | Policy and Regulation | 2018 |
| Handbook on Data Protection in Humanitarian Action | Manual/Guidelines | Brussels Privacy Hub (VUB) and International Committee of the Red Cross (ICRC) | Non-Governmental Organizations | Global | Humanitarian | 2020 |
| Policy on the Protection of Personal Data of Persons of Concern to UNHCR | Policy | UNHCR | Intergovernmental Institutions | Global | Humanitarian | 2018 |
| OCHA data responsibility guidelines 2021 | Manual/Guidelines | UN Office for the Coordination of Humanitarian Affairs (UN OCHA) | Non-Governmental Organizations | Global | Humanitarian | 2021 |
| Procedures for Ethical Standards in Research, Evaluation, Data Collection and Analysis | Manual/Guidelines | UNICEF | Intergovernmental Institutions | Global | Humanitarian | 2015 |
| Personal Data Protection and Privacy Principles | Principles | UN High-Level Committee on Management (HLCM) | Intergovernmental Institutions | Global | Not One Specific Sector | 2018 |
| Considerations for Using Data Responsibly | Policy | United States Agency for International Development (USAID) | Intergovernmental Institutions | Global | Research and Development | 2019 |
| The Geneva Declaration on Targeted Surveillance and Human Rights | Charter | Access Now and Government of Catalonia | International Independent Coalitions | International | Not One Specific Sector | 2022 |
| Recommendation of the Council concerning Guidelines Governing the Protection of Privacy and Transborder Flows of Personal Data | Manual/Guidelines | OECD (Organisation for Economic Co-operation and Development) | Intergovernmental Institutions | Global | Not One Specific Sector | 2013 |
| WHO Data Principles | Principles | UN - World Health Organization | Intergovernmental Institutions | Global | Health | 2020 |
| EU Data Governance Act (2021) | Law | European Union | Intergovernmental Institutions | Regional | Policy and Regulation | 2022 |
| UN Global Pulse Principles on Data Protection and Privacy | Manual/Guidelines | UNGP Global Data Access Initiative (GDAI) | Intergovernmental Institutions | Global | Research and Development | 2020 |
| Data Principles for UNDP | Principles | UNDP | Intergovernmental Institutions | Global | Research and Development | 2020 |
| UNICEF & UNFPA Policy on Personal Data Protection. | Policy | UNICEF | Intergovernmental Institutions | Global | Humanitarian | 2020 |
| UNICEF Data Quality Framework | Conceptual Framework | UNICEF | Intergovernmental Institutions | Global | Humanitarian | 2021 |
| WFP Guide to Personal Data Protection and Privacy | Manual/Guidelines | WFP | Intergovernmental Institutions | Global | Humanitarian | 2016 |
| ICAO Aeronautical Information Services Manual - Doc 8126 | Regulation Framework | ICAO - International Civil Aviation Organization | Intergovernmental Institutions | Global | Mobility | 2021 |
| An operational Data Governance Framework for New Zealand Government | Regulation Framework | Stats NZ | Governmental Institutions | Local | Policy and Regulation | 2019 |
| California Consumer Privacy Act (CCPA) | Law | California | Governmental Institutions | Local | Policy and Regulation | 2018 |
| African Union Data Policy Framework | Policy | African Union Commission | Intergovernmental Institutions | Regional | Policy and Regulation | 2022 |
| Japanese Basic Act on the Advancement of Public and Private Sector Data Utilisation | Law | Japan | Governmental Institutions | National | Policy and Regulation | 2016 |
| National Data Governance Framework: Information Systems for Health | Conceptual Framework | PAHO | Intergovernmental Institutions | Regional | Health | 2021 |
| Recommendation of the Council on Health Data Governance | Conceptual Framework | OECD | Intergovernmental Institutions | Global | Health | 2016 |
| IGAD Regional Health Data Sharing and Protection Policy FRAMEWORK | Policy | IGAD (Intergovernmental Authority on Development) | Intergovernmental Institutions | Regional | Health | 2022 |
| Technical Report D4.1 - Framework for security, privacy, risk and governance in data processing and management | Manual/Guidelines | ITU - International Telecommunication Union | Intergovernmental Institutions | Global | Telecommunications | 2019 |
| GSMA Guidelines on mobile money data protection | Manual/Guidelines | GSMA | International Independent Coalitions | Global | Telecommunications | 2018 |
| OPERATIONAL GUIDANCE DATA RESPONSIBILITY IN HUMANITARIAN ACTION | Manual/Guidelines | Inter-Agency Standing Committee (IASC) - (UN) | Intergovernmental Institutions | Global | Humanitarian | 2021 |
| ASEAN Data Management Framework | Manual/Guidelines | Association of Southeast Asian Nations (ASEAN) | Intergovernmental Institutions | Regional | Not One Specific Sector | 2021 |
| United Nations Fundamental Principles of Offical Statistics, Implementation Guidelines | Manual/Guidelines | United Nations Statistics Division (UNSTATS) | Intergovernmental Institutions | Global | Not One Specific Sector | 2015 |
| OECD AI principles | Principles | OECD | Intergovernmental Institutions | Global | Not One Specific Sector | 2019 |
| UNESCO Recommendation on the Ethics of AI | Conceptual Framework | UNESCO | Intergovernmental Institutions | Global | Other | 2021 |
| UNESCO's Internet Universality Indicators | Conceptual Framework | UNESCO | Intergovernmental Institutions | Global | Other | 2018 |
| Data Security Law of the People's Republic of China | Law | Republic of China | Governmental Institutions | Local | Not One Specific Sector | 2021 |
| Cybersecurity Law of the People's Republic of China | Law | Republic of China | Governmental Institutions | Local | Not One Specific Sector | 2017 |
| Recommendation of the OECD council on health data governance | Conceptual Framework | OECD | Intergovernmental Institutions | International | Health | 2016 |
| Inter-American Open Data Program to Combat Corruption - PIDA | Regulation Framework | OAS | Intergovernmental Institutions | Regional | Other | 2019 |
| APEC SECRETARIAT Personal Data Protection Policy | Policy | Asia-Pacific Economic Cooperation (APEC) | Intergovernmental Institutions | Regional | Policy and Regulation | 2021 |
| Canada's Personal Information Protection and Electronic Documents Act (PIPEDA) | Law | Government of Canada | Governmental Institutions | Local | Policy and Regulation | 2015 |
| UN Integrated Geospatial Information Framework (UN-IGIF) | Conceptual Framework | UN-IGIF | Intergovernmental Institutions | International | Other | 2019 |
| The Gemini Principles | Conceptual Framework | Centre for Digital Built Britain | Academia and Research Institutions | Global | Research and Development | 2018 |
| CARE Principles for Indigenous Data Governance | Conceptual Framework | Global Indigenous Data Alliance | International Independent Coalitions | Global | Other | 2019 |
| Health Data Governance Principles | Principles | Transform Health | International Independent Coalitions | Global | Health | 2022 |
| Cross-Border Data Policy Principles | Principles | Global Data Alliance | International Independent Coalitions | Global | Economics and Finance | 2021 |

## Appendix B: Extended Public Repository

At this link can be found an extended version of our repository, which served as the table of analysis for this research: https://docs.google.com/spreadsheets/d/1O6iPENGkQ6DLD3SQmZt3qIqGGnwmAeQbUtviqy5TYLk/edit?usp=sharing.